\documentclass[showpacs,amsmath,amssymb,floatfix,preprint,prb]{revtex4}

\usepackage{graphicx}
\usepackage{dcolumn}
\usepackage{bm}

\begin{document}

\title{Studying single-electron transistors by microwave and far-infrared absorption: theoretical results and experimental proposal}

\author{Ioan B\^aldea}
\email{ioan.baldea@pci.uni-heidelberg.de}
\altaffiliation[Also at ]{National Institute for Lasers, Plasma, 
and Radiation Physics, ISS, POB MG-23, RO 077125 Bucharest, Romania.}
\author{Horst K\"oppel} 
\affiliation{Theoretische Chemie,
Physikalisch-Chemisches Institut, Universit\"{a}t Heidelberg, Im
Neuenheimer Feld 229, D-69120 Heidelberg, Germany}

\begin{abstract}
We present theoretical results on microwave and far-infrared (FIR) absorption 
of single-electron transistors obtained within exact numerical diagonalization for 
finite clusters. They show that both the microwave and the FIR spectra consist 
of two maxima, whose origin can be understood physically. 
Our results on microwave absorption provide a physically intuitive qualitative interpretation 
of the Kondo splitting observed by Kogan et al [Science {\bf 304}, 1293 (2004)].
The present results 
on the FIR absorption supplement and provide a physical insight into previous 
results obtained by means of the numerical renormalization group.
Based on our theoretical results, we propose to conduct 
FIR experiments to determine the charging energy and other relevant parameters. 
\end{abstract}

{ 

\maketitle
\section{Introduction}
\label{sec-introduction}
In a single-electron transistor (SET), 
which consists of a quantum dot (QD) attached to two electrodes,
a small source-drain voltage yields a current flowing only for 
certain values of the gate potential $V_g$. 
\cite{Goldhaber-GordonNature:98,Goldhaber-GordonPRL:98,Wiel:00}
At temperatures below the Kondo temperature ($T<T_K$), 
conduction occurs in a $V_g$-range delimited by the 
situations where the energy of the dot level  $\varepsilon_d$
is such that 
the lowest Hubbard ``band'' or the highest Hubbard ``band'' 
are nearly resonant with the electrode Fermi energy $\varepsilon_F$,
$\varepsilon_d \approx \varepsilon_F$ and 
$\varepsilon_d + U \approx \varepsilon_F$, respectively.
Here, $U$ represents the dot charging energy, i.~e., the energy 
required to add an extra electron on the dot.
The zero-bias conductance $G$ reaches the unitary limit
$G = G_0 = 2e^2/h$ within the Kondo plateau, which is 
defined by $\varepsilon_F - U \alt \varepsilon_d \alt \varepsilon_F$.
\par
The charging energy represents a key parameter for SETs. 
The unpleasant fact is that 
in dc transport measurements of the zero-bias conductance $U$ cannot be directly determined, 
because the dot energy $\varepsilon_d$ 
cannot be directly controlled, but rather indirectly
via the potential of a ``plunger'' gate potential $V_g$, 
on which it linearly depends:
$ \varepsilon_d = \alpha V_g + \mbox{const.} $
If $V_{g,l}$ and $V_{g,u}$ denote the gate potentials whereat the lower 
and upper Hubbard bands 
become resonant, one gets $U = \alpha (V_{g,l} - V_{g,u})$. 
So, to determine $U$, in addition to the difference $V_{g,l} - V_{g,u}$, which is available
from the transport data, supplementary hypotheses are needed to  
deduce the conversion factor $\alpha$, which, although physically 
plausible, cannot be fully justified at the nanoscale 
and require assumptions or arguable extrapolations 
of macroscopic relations to the nanoscale. One way is to assume a certain phenomenological 
$T$-dependence (convolution of a Lorentzian with the derivative 
of the Fermi function) and to fit the width of the 
Coulomb blockade peaks $G(T)$.\cite{Goldhaber-GordonPRL:98,Amasha:05}
Another possibility is to resort to the capacitance model,
which describes the SET in terms of three effective capacities $C_g$, $C_s$, and $C_d$ between 
the dot and the gate, source, and drain, respectively.
\cite{AverinLikharev:86,KouwenhovenAustingTarucha:01,Kubatkin:03}
The conversion factor, expressed by $\alpha = C_g/C$ ($C\equiv C_g + C_s + C_d$), can then 
be obtained from the Coulomb diamonds of the stability diagram.
Even without inquiring whether such assumptions are justified, 
the inaccuracies of the parameters estimated in this way 
are rather large; uncertainties can be as large as $\sim 20$\%.\cite{Liang:02}
Therefore, utilizing more accurate or at least alternative methods of investigation
is highly desirable.
\par
It is a main goal of this paper to show that the far-infrared (FIR) absorption 
represents a possible alternative technique for the characterization 
of SETs and how such experiments can be conducted.
\par
The remaining part of this paper is organized in the following manner. 
In Sect.~\ref{sec-model} we expose the theoretical framework, and in 
Sect.~\ref{sec-comp} present all relevant computational details. 
In Sect.~\ref{sec-results}, exact numerical results for the full ac absorption spectra 
are presented and analyzed in terms of a few physically relevant many-electron states.
Sect.~\ref{sec-finite-size} is devoted to finite size effects. Next we discuss the two 
distinct spectral ranges significant for SETs separately: the microwave/radiofrequency 
absorption in Sect.~\ref{sec-rf} and the FIR absorption in Sect.~\ref{sec-fir}.
Experimental implications of the theoretical results for the FIR absorption are 
presented in Sect.~\ref{sec-exp}. Sect.~\ref{sec-conclusion} is devoted to conclusions.
\section{Theoretical framework}
\label{sec-model}
Following the usual procedure, we shall describe the SET within the 
Anderson single-impurity model
\cite{AverinLikharev:86,Ng:88,Glazman:88,Izumida:98}
\begin{eqnarray}  
H & = & 
\varepsilon_{F} \sum_{\sigma, n=-1}^{-M_L} a_{n,\sigma}^{\dagger} a_{n,\sigma}^{}
+ \varepsilon_{F} \sum_{\sigma, n=1}^{M_R}  a_{n,\sigma}^{\dagger} a_{n,\sigma}^{}
\nonumber \\ 
& &
- t \sum_{\sigma,n=-1}^{-M_L+1} \left(
  a_{n,\sigma}^{\dagger} a_{n-1,\sigma}^{} 
+ h.c. \right) \nonumber \\
& &
-t \sum_{\sigma,n=1}^{M_R-1} \left(
  a_{n,\sigma}^{\dagger} a_{n+1,\sigma}^{} 
+ h.c. \right) \label{eq-hamiltonian} \\
& & 
- t_{d} \sum_{\sigma} \left(a_{-1,\sigma}^{\dagger} d_{\sigma}^{} +  a_{+1,\sigma}^{\dagger} d_{\sigma}^{} 
+ h.c. \right) 
\nonumber \\
& &
+ \varepsilon_{d}  \sum_{\sigma} d_{\sigma}^{\dagger} d_{\sigma}^{} 
+  U d_{\uparrow}^{\dagger} d_{\uparrow}^{} d_{\downarrow}^{\dagger} d_{\downarrow}^{} . 
\nonumber
\end{eqnarray}  
The left ($L$) and right ($R$) electrodes are assumed to contain noninteracting electrons, 
which are characterized by the same bandwidth $D=4t$ and the same coupling $t_d$ to the dot.
The dot is modeled by a single level, whose energy $\varepsilon_d$ can be tuned by means of a gate potential,
as discussed in Sect.~\ref{sec-introduction}.
$a_{n,\sigma}$ ($a^{\dagger}_{n,\sigma}$) are 
annihilation (creation) operators for electrons in the left and right 
leads ($L, R$) and  
$d_{\sigma} \equiv a_{0,\sigma}$ ($d^{\dagger}_{\sigma}  \equiv a^{\dagger}_{0,\sigma}$) 
destroys (creates) electrons in the QD. The number of electrons will be assumed to be equal to 
the number of sites, $N=M_L + M_R + 1$.
\par
The quantity of interest, the frequency-dependent absorption coefficient 
$\mu(\omega)$ in the ground state $\Psi_0$ (case of zero temperature),
can be expressed as a sum of contributions of various excited 
states $\Psi_{\lambda}$ ($H \Psi_{\lambda} = E_{\lambda} \Psi_{\lambda}$)
\begin{equation}
\mu(\omega) = \omega  \sum_{\lambda \neq 0}
\vert\langle \Psi_{\lambda}\vert P_d\vert \Psi_0 \rangle\vert^{2} 
\delta\left(\omega - E_{\lambda} + E_0 \right),
\label{eq:alpha}
\end{equation} 
where $P_d$ is the QD dipole moment. Eq.\ (\ref{eq:alpha}) represents the 
result of the linear response theory by considering 
an ac electromagnetic perturbation $H^{\prime} = - P_d \mathcal{E}_0 \cos \omega t$.
Various aspects of the problem of a SET in an ac field
within the linear response approximation 
were previously considered in several studies (see, e.~g., 
Refs.~\onlinecite{CampoOliviera:03,Sindel:05,Laakso:08}).
Because the definition of the dipole operator $P_d$ for a point-like QD poses 
some problems, it is more convenient to express 
the matrix elements entering  Eq.\ (\ref{eq:alpha})
$\langle \Psi_{\lambda} \vert P_d \vert \Psi_0 \rangle = - i \hbar 
\langle \Psi_{\lambda} \vert j_d \vert \Psi_0 \rangle /(E_{\lambda} - E_0)$ 
in terms of the current operator $j_d = (j_{-1/2} + j_{1/2})/2$, as done
in similar cases,\cite{MeneghettiDipoleCurrent}
which can be unambiguously defined as 
$j_{n + 1/2} = i t_n (e/\hbar) \sum_{\sigma} (a_{n,\sigma}^{\dagger}  a_{n+1,\sigma} - \mbox{H.c.})$.
So, the ac absorption is specified by the spectral lines $\lambda $ characterized by the absorption intensities
$\mu_{\lambda}$ and the absorption frequencies $\omega_{\lambda}$ defined by
\begin{eqnarray}
\mu_{\lambda} & = & 
\frac{1}{\omega_{\lambda}}\vert\langle \Psi_{\lambda}\vert \hat{\tau}\vert \Psi_0 \rangle\vert^{2} \nonumber \\ 
\hat{\tau} & \equiv &  i 
\sum_{\sigma} \left( a_{-1,\sigma}^{\dagger} d_{\sigma} - a_{+1,\sigma}^{\dagger} d_{\sigma} - h.c.\right) 
\label{eq:mu} \\
\omega_{\lambda} & \equiv & E_{\lambda} - E_0 \ . \nonumber
\end{eqnarray} 
In the presentation and the discussion of the results on ac absorption, unless otherwise 
specified, we shall refer throughout to a SET in the Kondo regime 
($\varepsilon_F - U \alt \varepsilon_d \alt \varepsilon_F $).
Moreover, we can restrict ourselves to the range 
$\varepsilon_F - U/2 \alt \varepsilon_d \alt \varepsilon_F$
because of the particle-hole symmetry.
\section{Computational details}
\label{sec-comp}
Below, we shall present results on the ac absorption of a SET obtained by 
exact (Lanczos) numerical diagonalization. The method of computation we employ 
here is that used in our earlier works; see, e.~g., 
Refs.~\onlinecite{koeppel:84,Baldea:97,Baldea:2000,Baldea:2001a,Baldea:2002,Baldea:2004b,Baldea:2007,Baldea:2008,Baldea:2009a,Baldea:2009b}. Because the full details on this method were not published and because of the 
significant differences between our Lanczos implementation to compute the linear response 
and the more familiar continued fraction algorithm,\cite{HHK:80,fulde:91,dagotto:94} 
we describe it below for the benefit of the reader. 

In the first run, the Lanczos procedure is iterated until, after $\mathcal{N_L}$ iterations,
the lowest (ground state) energy $E_0$ converges. In the second run, by carrying out again 
$\mathcal{N_L}$ iterations and with the same starting Lanczos vector, 
the corresponding Ritz vector $\Psi_0$ 
is computed by accumulation without the need of storing the Lanczos vectors.  
To check that this vector represents indeed the accurately evaluated ground state $\Psi_0$, 
we straightforwardly compute the dispersion 
$\langle \Psi_0\vert (H - E_0)^2\vert \Psi_0\rangle ^{1/2}$ and convince ourselves that 
it is much smaller (usually $5-6$ orders of magnitude) than the lowest excitation energy.
The above scheme can also be used to reliably compute several lower excited eigenstates, 
but it is usually unpractical to target all the eigenstates $\Psi_{\lambda}$ needed to compute the 
linear response via Eq.~(\ref{eq:alpha}), e.~g., by orthogonalization on eigenvectors already converged 
in previous runs. The  reason is that many eigenvectors, which are not 
important for the linear response, are also targeted.
To ensure that the important eigenvectors are targeted, in a third Lanczos run, 
we employ a starting Lanczos vector adequate for the specific linear 
response considered. This is, in the present case, the normalized vector $P_d\vert\Psi_0\rangle$.
The needed matrix elements $\langle \Psi_{\lambda}\vert P_d \vert \Psi_0 \rangle$ 
are given by the first component of the tridiagonal vectors obtained in this third run.
Usually, a number of iterations comparable to $\mathcal{N_L}$ suffices for the third run. 
As an important test of the results for the linear response computed in this way, we 
always check whether they satisfy the sum rule, which can be deduced exactly from 
Eq.~(\ref{eq:alpha})
\begin{equation}
\sum_{\lambda} \vert\langle \Psi_{\lambda}\vert P_d\vert \Psi_0 \rangle\vert^{2} 
=  \langle \Psi_0\vert P_d^2\vert \Psi_0 \rangle \ ,
\label{eq:sum-rule}
\end{equation} 
because the r.h.s. is known, namely the squared norm of the vector 
$P_d\vert\Psi_0\rangle$.
In certain cases, the linear response computed within the third run does not satisfy the 
above sum rule, e.~g., 
because of spurious vector duplication. Therefore, to be always on the safe side, 
we carry out a fourth Lanczos run, wherein,
similar to the second run, we also compute and store all those Ritz vectors 
$\Psi_{\lambda}$, which where 
found to have a significant spectral weight  
$\vert\langle \Psi_{\lambda}\vert P_d \vert \Psi_0 \rangle \vert^2$ 
[in practice, above $10^{-5}$ of the r.h.s.\ of (Eq.~(\ref{eq:sum-rule})]
in the third run.
Storing these vectors $\Psi_{\lambda}$ is not much more demanding
than storing the ground state $\Psi_0$ alone, 
because for all the problems we investigated so far, at most $\sim 10 - 20$ 
Ritz vectors are important. 
The real, prohibitive limitation remains, as in all exact diagonalization approaches, 
the cluster size.
Again, we check that these Ritz vectors 
are accurate eigenvectors by straightforwardly computing the dispersions 
$\langle \Psi_{\lambda}\vert (H - E_{\lambda})^2\vert \Psi_{\lambda}\rangle ^{1/2}$.
By using these eigenvectors $ \Psi_{\lambda}$ we finally compute the linear response from 
Eq.~(\ref{eq:alpha}) and convince ourselves that all important eigenvectors have been 
targeted by checking the sum rule (\ref{eq:sum-rule}). 

Proceeding in this way, 
the computing time is at most $\sim 1.5 - 2$ times larger than for implementations of 
the continuous fraction algorithm,\cite{HHK:80,fulde:91,dagotto:94} but 
we can safely rule out any numerical artefacts and have the guarantee 
that the solution obtained is mathematically exact. In addition and equally important, 
this method allows us to 
compute and resolve \emph{individual} nearly degenerate spectral lines, a 
situation where the information that can be extracted from convoluted spectra 
provided by the continued fraction algorithm does not suffice.
This represents a quite relevant aspect 
for SETs and other QD-based nanosystems, where nearly degenerate states with the same symmetry 
(avoided crossings) are often encountered; 
see Refs.~\onlinecite{Baldea:2008,Baldea:2009a,Baldea:2009b,Baldea:2010a} and 
Sect.~\ref{sec-rf}.
\section{Exact results on the full ac absorption spectra and their physical interpretation}
\label{sec-results}
Numerical exact results for frequencies and intensities of all the ac absorption signals 
obtained as described in Sect.~\ref{sec-comp} are collected in Fig.~\ref{fig:abs}. 
They have been obtained for $N=11$ 
and parameter values, which are typical for real cases: $t = 0.5$\,eV 
(electrode bandwidth $D = 4t = 2$\,eV), $t_d = 0.2$\,meV, and $U = 8$\,meV. 
We emphasize that these are numerical exact results, obtained by using all the 213444 
multielectronic configurations of the eleven-site cluster with eleven electrons 
and a total spin projection $S_z=1/2$. Based on the considerations of Sect.~\ref{sec-comp}
we can safely state that the ac spectrum of the investigated cluster solely consists of 
four relevant absorption signals. The other transitions, although allowed 
by symmetry, are completely irrelevant, as their intensities are orders of 
magnitude smaller and are therefore invisible in Fig.~\ref{fig:abs}b.\cite{dot-spectra}

Exact results on the SET ac absorption have of course their own importance, but 
do not yet provide much physical insight into the problem. Since the above 
exact results show that only four optical transitions are important, one can expect that,
out of numerous multielectronic configurations 
(namely, $213444$, see above, in the case under consideration,
of an eleven-site cluster with a total spin projection $S_z=1/2$),
there should only exist a few many-body states, which are relevant. If so, the problem 
is of course to identify them and to unravel their physical content. 
This shall be done next.

\begin{figure}[htb]
\centerline{
\includegraphics[width=0.38\textwidth,angle=-90]{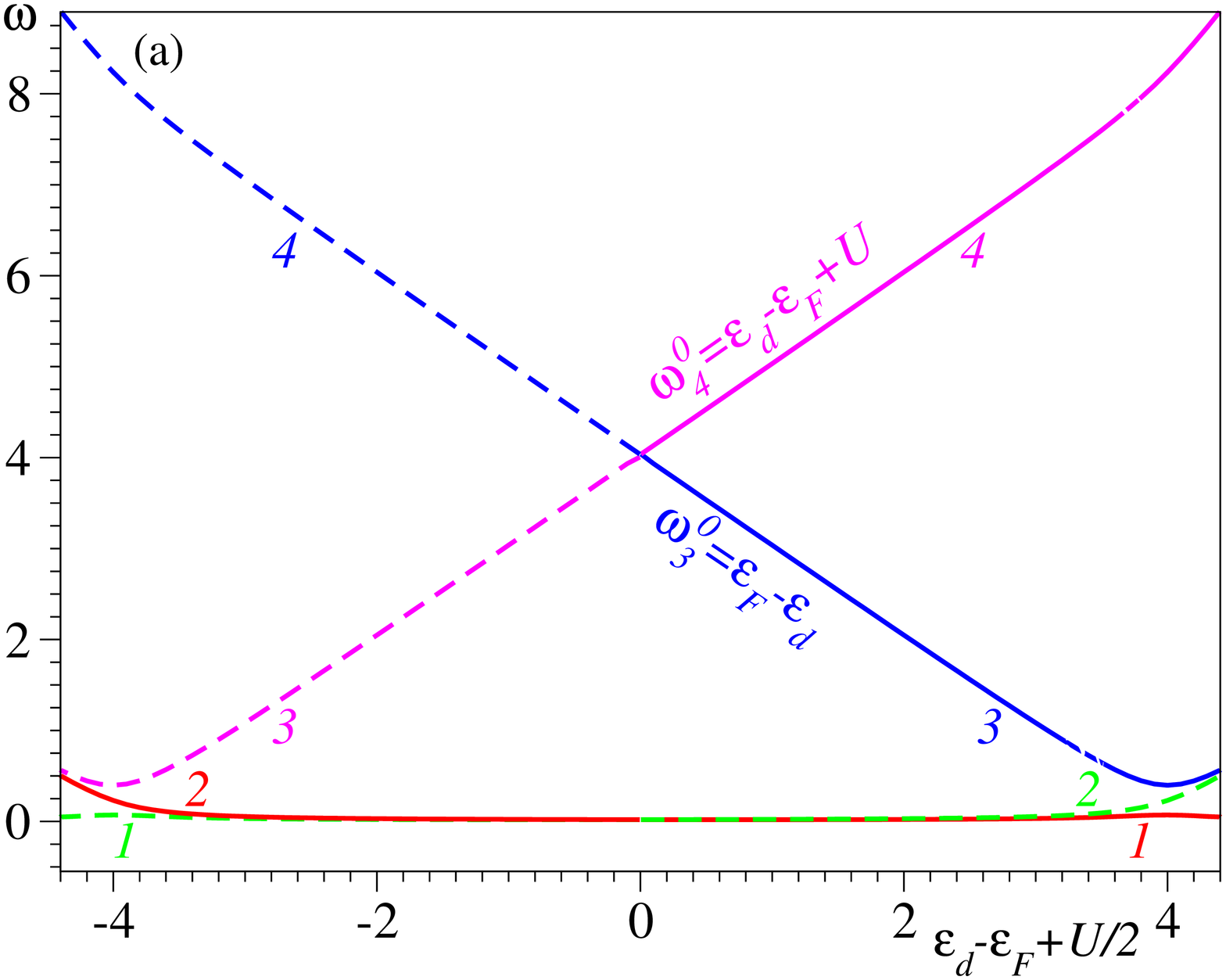}
\includegraphics[width=0.38\textwidth,angle=-90]{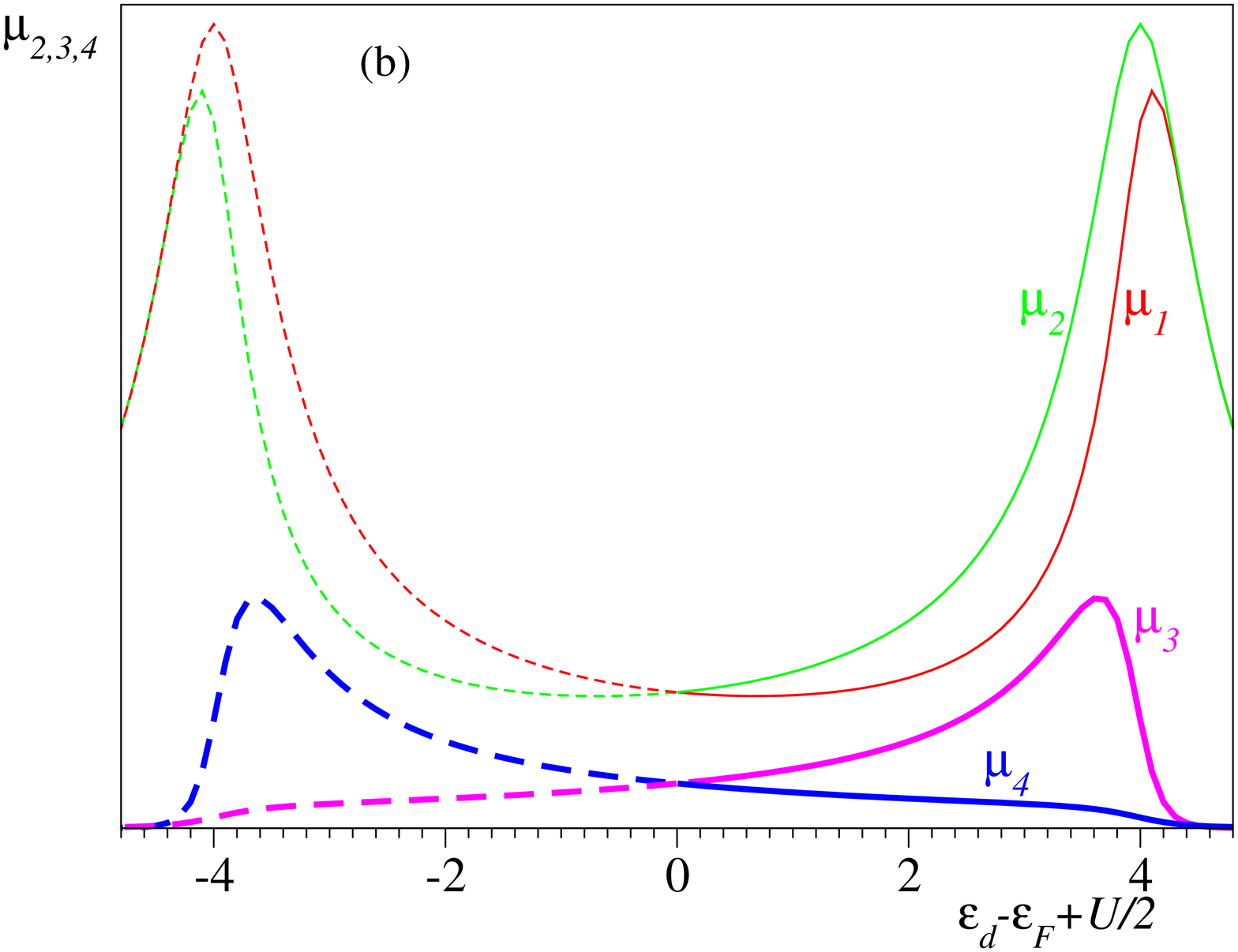}}
\caption{\label{fig:abs} (Color online) Dependence on the dot energy $\varepsilon_d$ of 
the four relevant (a) absorption frequencies $\omega$ (in meV) 
and (b) absorption intensities 
$\mu$ (in arbitrary units) of the optical transitions with significant spectral intensities.
In (a), deeper within the Kondo plateau, the exact frequencies $\omega_{3,4}$ 
are well approximated by $\omega_{3,4}^0$. The dashed lines are the analytical continuations 
of the full lines. 
The parameter values are: $t=0.5$\,eV, $t_d=0.2$\,meV and 
$U=8$\,meV.}
\end{figure}

There are nine such significant many-electron states.
These configurations ($\vert 1 \rangle$ to $\vert 9 \rangle$) 
are schematically 
shown in Fig.~\ref{fig:states}. Configurations $\vert 1 \rangle $ to
$\vert 5 \rangle $ correspond to one electron on the dot, in $\vert 6 \rangle $ and 
$\vert 7 \rangle $ the dot level is vacant, while in $\vert 8 \rangle $ and $\vert 9 \rangle $
it is occupied by two electrons. 
A superficial glance at the schematic representation of Fig.~\ref{fig:states} can easily 
overlook both the underlying physics and the computational effort involved, 
and therefore a comment 
is in order at this point. 
Out of the electrons in the two electrodes, only those occupying the Fermi levels are shown 
for the nine states of Fig.~\ref{fig:states}.
For these nine states, the single-particle states of the electrons in the 
electrodes are in momentum ($k$) space, and 
not in the real (site, $n$) space, in which the exact numerical diagonalization is carried out
because the Hamiltonian matrix, Eq.~(\ref{eq-hamiltonian}), is sparse.
A single-particle $k$-state, e.~g., in the left electrode represents a superposition 
of $M_L$ single-particle $n$-states. In addition, one should note that the electrons in electrodes
depicted in Fig.~\ref{fig:states} represent electrons at the Fermi level. This means that 
these electrons are 
delocalized over the electrodes. Consequently, although we show below that 
the approximative description in terms of the nine relevant states is accurate,
it is not a priori obvious that the problem can be reduced or reasonably approximated 
by studying a three-site cluster. To summarize, each of the nine states depicted in 
Fig.~\ref{fig:states} contains in fact numerous multielectronic configurations in the real space. 
However, what is physically important is the existence of a very reduced number of the 
relevant states.

\begin{figure}[htb]
\includegraphics[width=0.38\textwidth,angle=-90]{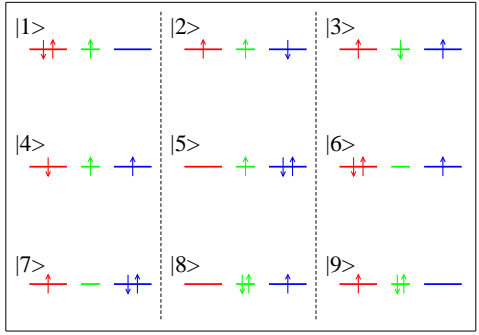}
\caption{\label{fig:states} (Color online) Multielectronic 
configurations with significant contributions to the ground state 
$\Psi_0$ and the excited 
states $\Psi_{1,2,3,4}$ 
important for ac absorption. For each configuration, we show the 
electrons at the Fermi levels of the left and right electrodes, and on the dot 
(red, blue, and green, respectively). In either electrode, the single-electron states 
below the Fermi level are occupied.}
\end{figure}

The discussion below proceeds in terms of these nine many-body states with significant 
contributions to the ground state and the 
four excited states 1, 2, 3, and 4 depicted in Fig.\ \ref{fig:abs}.
From these nine most relevant states one can construct the following 
states with definite spin $S_z = S = 1/2$ (notice that the total electron number is odd), which are 
either even ($g$) or odd ($u$) under space inversion
\begin{eqnarray}  
\vert u_1 \rangle & = & 
\left( \vert 1 \rangle - \vert 5 \rangle + \vert 2 \rangle + \vert 4 \rangle - 
2\ \vert 3 \rangle \right) / \sqrt{8} ; \nonumber \\
\vert u_2 \rangle & = &  \left( \vert 6 \rangle - \vert 7 \rangle \right)/\sqrt{2} ; \nonumber \\
\vert u_3 \rangle & = & \left( \vert 8 \rangle - \vert 9 \rangle \right)/\sqrt{2} ; \nonumber \\
\vert g_1 \rangle & = & \left( \vert 1 \rangle + \vert 5 \rangle - \vert 2 \rangle + \vert 4 \rangle \right)/2 ; \label{eq-g_u} \\
\vert g_2 \rangle & = & \left( \vert 1 \rangle + \vert 5 \rangle + \vert 2 \rangle - \vert 4 \rangle \right)/2 ; \nonumber \\
\vert g_3 \rangle & = & \left( \vert 6 \rangle + \vert 7 \rangle \right)/\sqrt{2} ; \nonumber \\
\vert g_4 \rangle & = & \left( \vert 8 \rangle + \vert 9 \rangle \right)/\sqrt{2} . \nonumber
\end{eqnarray}  
The eigenstates important for ac absorption can be well approximated as
\begin{eqnarray} 
\left \vert \Psi_0 \right \rangle & \simeq & 
\vert u_1 \rangle \cos \chi - \vert u_2 \rangle \sin \chi \to \vert u_1 \rangle ; \nonumber \\ 
\left \vert \Psi_1 \right \rangle & \simeq & 
\vert g_1 \rangle \cos \theta -   \vert g_3 \rangle \sin \theta \to \vert g_1 \rangle ;\nonumber \\
\left \vert \Psi_2 \right \rangle & \simeq & \vert g_2 \rangle ; \label{eq-Psi} \\ 
\left \vert \Psi_3 \right \rangle & \simeq & 
\vert g_1 \rangle \sin \theta +  \ \vert g_3 \rangle cos \theta \to \vert g_3 \rangle ; \nonumber \\
\left \vert \Psi_4 \right \rangle & \simeq & \vert g_4 \rangle . \nonumber
\end{eqnarray}  
Eqs.~(\ref{eq-Psi}) hold for  $\varepsilon_F - U/2 < \varepsilon_d < \varepsilon_F$. 
We can restrict ourselves to this range because of the particle-hole symmetry.
For $\varepsilon_F - U < \varepsilon_d < \varepsilon_F - U/2$, the states $\vert 6 \rangle $ and
$\vert 7 \rangle $ must be replaced by $\vert 9 \rangle $ and $\vert 8 \rangle $, 
and vice versa. 

To illustrate that the eigenstates $\Psi_{0,1,2,3,4}$ computed exactly are indeed 
very well approximated by the 
expressions in the r.h.s.\ of the symbols $\simeq$ in Eqs.~(\ref{eq-Psi}),
we present in Figs.~\ref{fig:w-psi-0} and \ref{fig:w-psi-exc} 
the curves of the weights 
$p_{1,2}^{0} \equiv \vert\langle u_{1,2}\vert \Psi_0\rangle\vert^2$ 
and
$p_{i}^{j} \equiv \vert\langle g_{j}\vert \Psi_{j}\rangle\vert^2$ ($i, j = 1,3$).
In all these cases, the two functions entering the 
expressions in the r.h.s.\ of Eqs.~(\ref{eq-Psi}) exhaust 
the expansions of the exact eigenstates $\Psi_{0,1,3}$ within an accuracy of 
$\sim 10^{-3}$. This fact fully justifies the use of the intuitive notations 
in terms of cosines and sines in Eqs.~(\ref{eq-Psi}), $\cos^2 \chi = p_1^0$, 
$\cos^2 \theta = p_1^1$. As concerns the other two exact eigenstates, 
the approximations $\vert \Psi_{2,4}\rangle \simeq \vert g_{2,4}\rangle$, 
are also accurate within $\sim 10^{-3}$.
\begin{figure}[htb]
\includegraphics[width=0.38\textwidth,angle=-90]{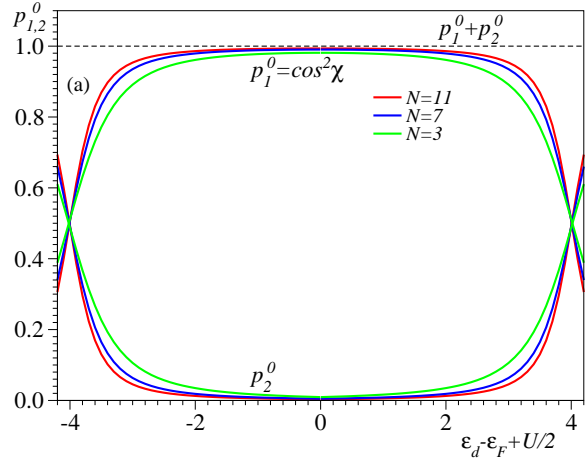}
\caption{\label{fig:w-psi-0} (Color online) $\varepsilon_d$-dependence  
of the weights $p_{1,2}^{0} \equiv \vert\langle u_{1,2}\vert \Psi_0\rangle\vert^2$ 
of the states $u_{1,2}$ entering the linear combination of Eq.~(\ref{eq-Psi}) for clusters with 
$N=3,7,11$ sites. Parameter values as in Fig.~\ref{fig:abs}.
For all $N$'s, the deviation from unity of the sum  $p_{1}^{0} + p_{2}^{0}$
(at most $\sim 10^{-3}$) 
is invisible within the drawing accuracy.}
\end{figure}
\begin{figure}[htb]
\includegraphics[width=0.38\textwidth,angle=-90]{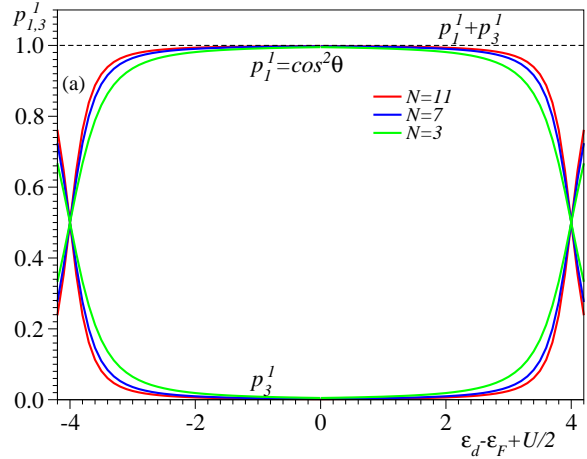}
\caption{\label{fig:w-psi-exc} (Color online) Curves for 
$p_{1}^{1} \equiv \vert\langle g_{1}\vert \Psi_1\rangle\vert^2$ and 
$p_{3}^{1} \equiv \vert\langle g_{3}\vert \Psi_1\rangle\vert^2$ 
similar to Fig.~\ref{fig:w-psi-0}. 
For all $N$'s, the $p_{1}^{1}$-curve 
cannot be distinguished within the drawing accuracy from that for 
$p_{3}^{3} \equiv \vert\langle g_{3}\vert \Psi_3\rangle\vert^2$, and the 
$p_{3}^{1}$-curve from that for  
$p_{1}^{3} \equiv \vert\langle g_{1}\vert \Psi_3\rangle\vert^2$.
For all $N$'s, the deviation from unity of the sum  $p_{1}^{1} + p_{3}^{1}$
(at most $\sim 10^{-3}$) 
is invisible within the drawing accuracy.}
\end{figure}
As visible in Figs.~\ref{fig:w-psi-0} and \ref{fig:w-psi-exc}, 
deeper within the Kondo regime, $cos\chi \simeq 1$ and $cos\theta \simeq 1$, 
and therefore 
$\vert \Psi_{0,1,3} \rangle $ are reasonably approximated as expressed in the r.h.s.\ of  
the arrows in Eqs.\ (\ref{eq-Psi}).
Bearing this in mind and inspecting Eqs.~(\ref{eq-Psi}) and (\ref{eq-g_u}) and Fig.~\ref{fig:states},
one can identify two groups of important eigenstates, which are well separated energetically. 
The first group comprises the eigenstates $\Psi_{0,1,2}$, which basically consist of superpositions 
of the nearly degenerate configurations $\vert 1 \rangle - \vert 5 \rangle $, 
corresponding to states with a singly occupied dot. This fact
nicely reveals the spin entanglement and the role of the coherent superpositions of all the 
possible spin flip processes 
($\vert 1 \rangle \rightleftharpoons \vert 3 \rangle$,
$\vert 3 \rangle \rightleftharpoons \vert 5 \rangle$,
$\vert 2 \rangle \rightleftharpoons \vert 3 \rangle$,
$\vert 4 \rangle \rightleftharpoons \vert 3 \rangle$)
in the formation of the nearly degenerate states $ \Psi_{0,1,2} $ 
important for the Kondo effect. The absorption frequencies $\omega_{1,2}$ of these 
optical transitions are low, falling into the microwave \cite{Kogan:04a} or even radiofrequency (rf) range.
The second group comprises the higher energy states $\Psi_3$ and $\Psi_4$, which 
correspond to a dot that is either doubly occupied or empty. Loosely speaking, they amount to 
excite a particle-hole pair, wherein the hole state is on the dot and the particle state
in electrodes, or \emph{vice versa}. The corresponding absorption frequencies,
$\omega_3 \simeq \omega_3^0 = \varepsilon_F - \varepsilon_d$
and 
$\omega_4 \simeq \omega_4^0 = \varepsilon_d + U - \varepsilon_F$ (cf.~Fig.~\ref{fig:abs}a),
are of the order of the charging energy $U$, falling therefore into the FIR range.
\section{Finite-size effects}
\label{sec-finite-size}
As is well known, the drastic limitation of the exact numerical diagonalization to rather small 
cluster sizes $N$ often precludes a reliable finite scaling analysis. There are well known examples 
(see, e.~g., Refs.~\onlinecite{Baldea:97,Baldea:99a,Baldea:2001b}) of non-monotonic $N$-dependent properties,
or qualitatively different behaviors at smaller and larger $N$ due to a different underlying 
physics (see, e.~g., Ref.~\onlinecite{Baldea:2001b})
at the sizes where exact numerical diagonalization is feasible. This limitation is even more severe in 
the case of SETs, in the sense that not even all these small sizes can be included in a 
finite-scale analysis. A careful selection of the $N$-values to be included in the finite-scale analysis 
is often necessary, as is well known, e.~g., in the case of cyclic polyenes C$_N$H$_N$ or related systems, 
where H\"uckel ($N=4n+2$) and anti-H\"uckel ($N=4n$) systems behave differently 
($n$ is an integer); see, e.~g., Refs.~\onlinecite{Baldea:99a,Baldea:99b,Baldea:2001a,Baldea:2001b} 
and references cited therein. 
With our implementation described in Sect.~\ref{sec-comp}, 
we can reliably treat the linear response of half-filled clusters up to 
$N=14$, amounting to a dimension of the Hilbert space of 11,778,624. 
This is not too much different 
from the largest size ($N=12$) of most recent studies on the dc-conductivity of model 
(\ref{eq-hamiltonian}).\cite{Heidrich:09} In view of the analysis in terms of the relevant 
many-body states of Fig.~\ref{fig:states}, 
it is clear that considering symmetric clusters (identical electrodes) is advantageous.
Because short electrodes with an even number of sites are known to yield spurious results 
(compare Ref.~\onlinecite{BuesserWrong:04} with Ref.~\onlinecite{Heidrich:09}), what remains
is to consider electrodes with an odd number of sites, which mimic ``metallic'' 
electrodes (i.~e., electrodes with a partially occupied Fermi level).\cite{Baldea:2008b,Baldea:2009a} Concretely, this means that 
we are left with the values $N=3, 7, 11$. Obviously, one cannot expect 
to reliably deduce a scaling law solely based on these three $N$-values.

In view of the aforementioned limitations, similar to our previous 
works,\cite{Baldea:2008b,Baldea:2009a} we shall simply inspect whether 
the relevant properties computed for $N=3, 7, 11$ are significantly  
size dependent or not. Typical results are shown in Figs.~\ref{fig:w-psi-0}, 
\ref{fig:w-psi-exc}, and \ref{fig:all-omega}. They reveal that certain quantities,
like the lowest excitation energies $\omega_{1,2}$ of Fig.~\ref{fig:all-omega}b 
are strongly size dependent. Obviously, such results for  $\omega_{1,2}$ 
of exact diagonalization cannot 
be trusted, at least not quantitatively (see also Sect.~\ref{sec-rf}). 
\begin{figure}[htb]
\centerline{
\includegraphics[width=0.38\textwidth,angle=-90]{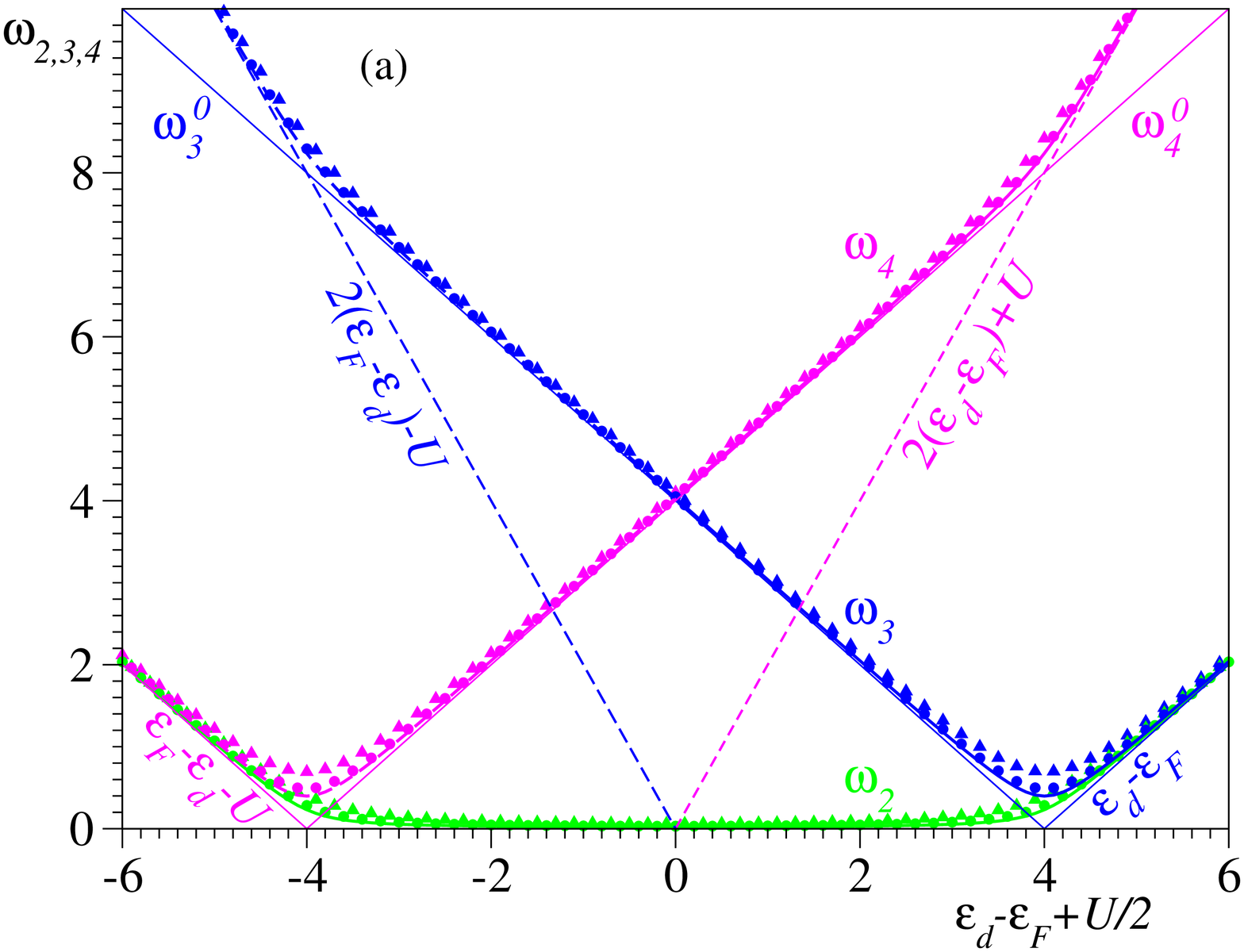}
\includegraphics[width=0.38\textwidth,angle=-90]{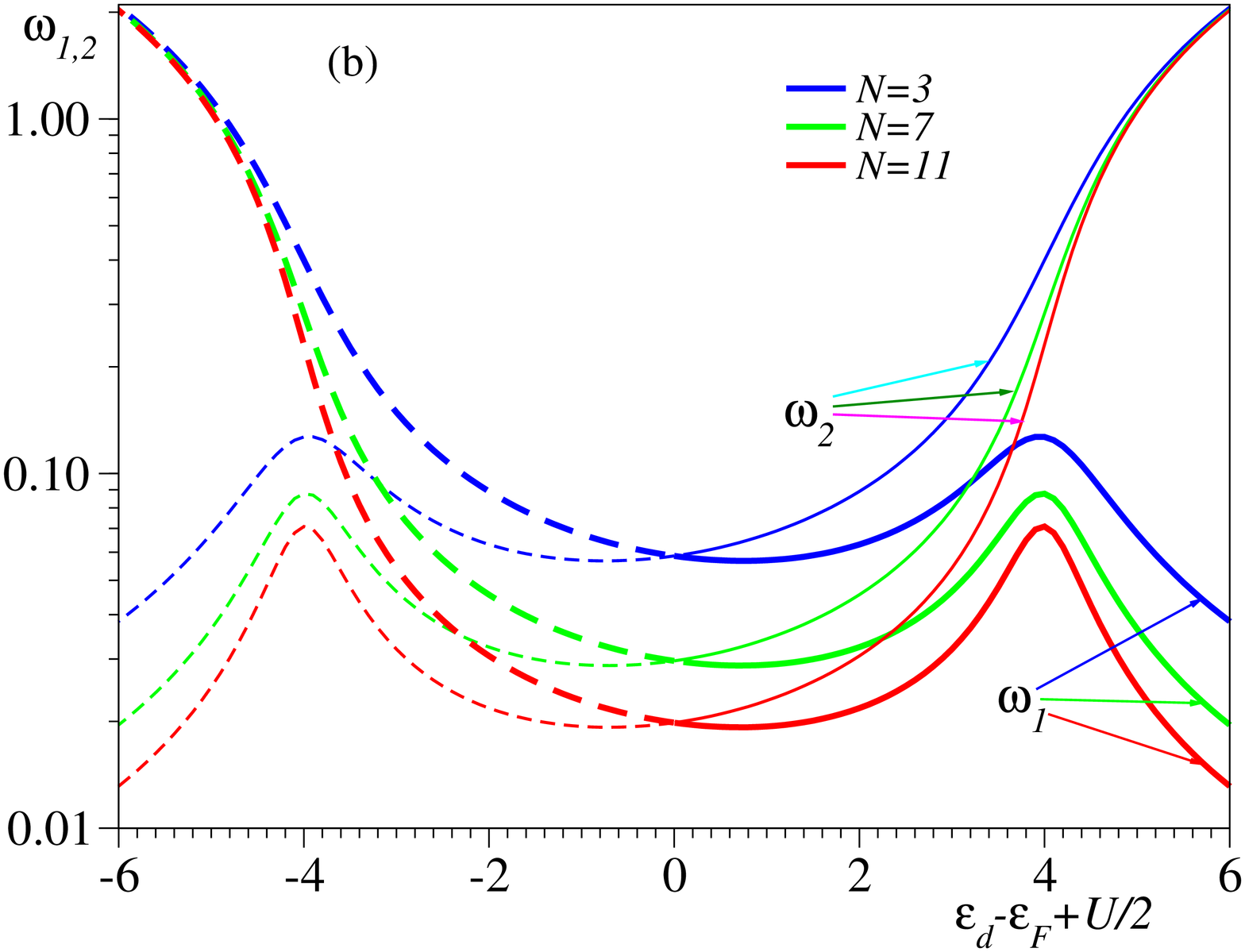}
}
\caption{\label{fig:all-omega} (Color online) Results on the 
(a) higher (FIR) frequency $\omega_{3,4}$ 
and (b) lower (rf/microwave) frequency $\omega_{1,2}$ ac absorption for 
several cluster sizes $N$
and same parameter values as in Fig.~\ref{fig:abs}. 
In panel (a), the triangles and circles are for clusters with $N=3$ 
and $N=7$, respectively, and solid lines are for clusters with $N=11$. 
The latter cannot be distinguished within the drawing accuracy 
from those of asymmetric clusters, wherein the dot is attached to a single 
electrode with 7, 9, 11, and 13 sites. In the Kondo regime, 
$\omega_{3,4}$ are only slightly 
size dependent, while $\omega_{1,2}$ are strongly size dependent. 
Notice the logarithmic scale on the ordinate in panel (b).}
\end{figure}

But, similarly to the examples presented in Refs.~\onlinecite{Baldea:2008b} and \onlinecite{Baldea:2009a}, 
there also exist quantities, which only slightly depend on $N$.
Most important for the main purpose of this work, 
this is the case of the two higher absorption frequencies
$\omega_{3,4}$ of Fig.~\ref{fig:all-omega}a, the key quantities to be measured in the 
FIR experiments we propose here (see Sect.~\ref{sec-exp}), 
Therefore, to give further support to the fact why we believe that, deeper in the Kondo plateau, 
the results for the curves $\omega_{3,4}(\varepsilon_d)$ are not significantly affected 
by finite size effects, we carried out supplementary calculations. Namely, we 
considered asymmetric 
clusters, wherein the QD is attached to the end of a single ``metallic'' 
electrode with an odd number of sites $N_u$. 
This procedure, which amounts to unfold the original symmetric 
cluster,\cite{CampoOliviera:03,Mehta:06,Heidrich:09}
has the advantage that the size $N_u$ of the single electrode can be larger, roughly 
twice that of one electrode of a symmetric cluster. The largest relevant (odd) size  
that we can treat by exact diagonalization is $N_u=13$ ($13+1$ sites). 
The shortcoming of the asymmetric cluster is that it misses 
the two lowest excitations $\omega_{1,2}$ related to the coherent spin fluctuations 
responsible for the Kondo effect. The first excitation of the asymmetric cluster, which is
almost degenerate with the singlet ground state, 
is a spin triplet, and the 
small singlet-triplet splitting could be considered as the counterpart of 
$\omega_{1,2}$ in symmetric clusters. However, this triplet excited state 
is irrelevant for the spin conserving ac absorption processes.
Most important is that the next two excitations of the asymmetric cluster 
are singlet states, which are optically active, and their 
energies are the counterpart of the above $\omega_{3,4}$. As noted in the caption of 
Fig.~\ref{fig:all-omega}a, the curves for $N_u=7,9,11,13$ cannot be distinguished from 
those of the symmetric cluster with $N=11$, which is the counterpart of the 
asymmetric cluster with $N_u=5$. 

For completeness, we mention that the size dependence of 
$\omega_{3,4}$ remains weak even beyond the Kondo plateau (cf.~Fig.~\ref{fig:all-omega}),
although this fact is not very important because of the small absorption intensities
(cf.~Fig.~\ref{fig:abs}b). There, the physical character of the $\omega_{3,4}$-excitations 
is different. Within the Kondo plateau they are related to excitations of a 
particle-hole pair, while beyond the mixed valence points they are related to 
excitations of two particle-hole pairs. This becomes clear if one inspects Fig.~\ref{fig:all-omega}a,
where the energies of the latter processes in the 
absence of electrode-dot coupling ($t_d \to 0$) are represented by the thin lines 
$2(\varepsilon_d -\varepsilon_F ) + U = (\varepsilon_d -\varepsilon_F ) + (\varepsilon_d + U -\varepsilon_F)$ 
and 
$2(\varepsilon_F -\varepsilon_d ) - U = 
(\varepsilon_F -\varepsilon_d ) + (\varepsilon_F -\varepsilon_d - U)$.\cite{1ph-2ph} 
A similar change in the physical character can be seen, e.~g., in the mixed valence 
region between the singly occupied and the vacant dot. There, the curve $\omega_3$,
which corresponds to the excitation of an electron from the singly occupied dot 
into electrodes ($\omega_3 \approx \omega_3^0 = \varepsilon_F - \varepsilon_d$), 
evolves into that amounting 
to bring an electron from electrodes to the vacant dot 
($\omega_3 \approx \varepsilon_d - \varepsilon_F$); 
see the lower right corner of Fig.~\ref{fig:all-omega}a.   
 
By inspecting Figs.~\ref{fig:w-psi-0} and \ref{fig:w-psi-exc}, one may argue that the 
size dependence of the wave functions $\Psi_0$, $\Psi_1$, and $\Psi_3$ is comparable;
so, where does the difference between the size dependence of $\omega_{1,2}$ on one side 
and $\omega_{3,4}$ on the other side come from? The reason is the following. 
While the $\omega_{1,2}$-values are close to $0$, the $\omega_{3,4}$-values vary 
close to the $\omega_{3,4}^{0}$-values.
which correspond to electrode-dot excitations in the limit of vanishing 
electrode-dot coupling ($t_d \to 0$), and are large ($\sim U$) 
deeper within the Kondo plateau. In fact, the size dependence of the difference 
$\omega_{3,4} - \omega_{3,4}^{0}$ is comparable to that of
$\omega_{1,2} - 0$, as seen in Fig.~\ref{fig:all-omega}a.
It is the same strong $N$-dependence of $\omega_{1,2}$ that makes $\mu_{1,2}$ 
[cf.~Eq.~(\ref{eq:mu})] strongly size dependent; the matrix elements 
of the hopping operator $\hat{\tau}$ are nearly $N$-independent. 

Although the results presented above have shown that the size dependence of the two higher 
optical transitions 
is not substantial, the important question is whether the absorption peaks $\mu_{3,4}$ 
survive when the cluster is linked to infinite electrodes. 
Based on our previous investigation of photoionization \cite{Baldea:2009a} 
and on extensive calculations of the FIR absorption in broad ranges of SET parameters,
we expect the following. As the size increases, the single-electron levels in electrodes
become more and more dense, and the straight lines $\omega_{3,4}^0$ of Figs.~\ref{fig:abs} or \ref{fig:all-omega}
will intersect the numerous horizontal (i.~e., $\varepsilon_d$-independent) 
lines corresponding to excitations of particle-hole pairs in electrodes,
in a way similar to that of the energies of various ionization processes 
(one-hole, two-hole--one-particle, \emph{etc}) shown in Fig.~2a of Ref.~\onlinecite{Baldea:2009a}.
Similar to Ref.~\onlinecite{Baldea:2009a}, this gives rise to a sequence of avoided crossings, but 
the spectral intensity remains concentrated in two diabatic states, as if these 
intersections were absent. From this perspective, one can also understand why, sufficiently away from 
the mixed-valence points, the curves for 
$\omega_{3,4}(\varepsilon_d)$ for the symmetric three-site cluster of Fig.~\ref{fig:all-omega}a 
represent reasonable approximations: roughly, they correspond to one electron-hole 
pair excitations, wherein the state of one mate of the pair is on the dot and the other 
at the electrode Fermi level. For the same reason, even the asymmetric two-site cluster
provides a qualitatively correct description of the FIR absorption.

As is well known,\cite{Datta:05} a weak electrode-dot coupling yields a small broadening 
($\Gamma \simeq 2 t_d^2/t$) of the isolated dot level $\varepsilon_d$.
The analysis of Sect.~\ref{sec-results} indicated that, basically, each of these transitions amounts to excite 
an electron-hole pair. Therefore, the electrode-dot coupling should reflect itself 
in a small broadening of the FIR peaks centered on the values 
$\omega_3 \simeq \omega_3^0 = \varepsilon_F - \varepsilon_d $ and 
$\omega_4 \simeq \omega_4^0 = \varepsilon_d + U - \varepsilon_F $,
which replace the delta-shaped $\mu_{3,4}$-lines of the finite cluster. 
From a strictly mathematical standpoint, to demonstrate that these FIR peaks survive when the finite cluster 
is linked to real electrodes,
we can simply invoke their presence in the numerical renormalization group (NRG) results,\cite{CampoOliviera:03} 
which are exact and consider infinite electrodes. 
\section{Radiofrequency/Microwave absorption}
\label{sec-rf}
The existence of two electromagnetic transitions 
$\vert \Psi_0 \rangle \to \vert \Psi_{1,2} \rangle $ 
with low absorption frequencies $\omega_{1,2}$ in the rf/microwave range 
is a remarkable theoretical result, 
because it is directly related to the 
recent experimental findings in SETs irradiated with microwaves.\cite{Kogan:04a,Wingreen:04}
Unfortunately, at present we cannot offer a reliable quantitative analysis and must 
restrict ourselves to a few qualitative considerations.
The first, obvious reason of this impossibility 
is the strong size dependence of the results
discussed in Sect.~\ref{sec-finite-size}. But there still exists another reason. 
As the electrodes become longer and longer ($N \to \infty$), we expect that  
$\omega_1$ tends to the width of the Kondo resonance $\sim T_K$.
At larger $U$, this width falls off exponentially with $U$, while our exact 
diagonalization data exhibit a much weaker, power law decrease with $U$. This $U$-dependence 
is similar to that of the width in the density of states obtained 
within a one-particle Green function approach.\cite{Chiappe:03} 
In that approach, also adopted in a series of other works 
(see Ref.~\onlinecite{Heidrich:09} and citations therein), 
the finite cluster is embedded into infinite electrodes via a Dyson equation, 
wherein the self-energy is 
supposed to be not affected by electron correlations. 
We are not aware of similar developments for the two-particle Green function needed 
to compute the ac absorption. 
Still, the aforementioned similar and (in this respect) 
incorrect $U$-dependence 
of that approach and the present one seems to
signal the need for a method that (presumably approximately but accurately enough) 
accounts for correlations in 
clusters of sizes much larger than the exact diagonalization can handle. 
In this sense, we think that the description of Sect.~\ref{sec-results} 
in terms of a few relevant many-body states is useful, 
since it emphasizes the similarity 
of the lowest two frequencies $\omega_{1,2}$ to a tunnel splitting.
The coherent spin fluctuations embodied into the functions $\Psi_{0,1,2}$ 
expressed by Eqs.~(\ref{eq-g_u}, \ref{eq-Psi}) amount to a coherent 
tunneling between configurations that are 
classically degenerate and have indeed similarities to the tunneling 
between the degenerate minima of a symmetric double well potential.
Most relevant, exponential decays of the 
tunnel splittings with the interaction strength are 
typical.\cite{Baldea:2001a,Baldea:2001b}
In view of the severe size limitation within exact numerical diagonalization,
and because it is unlikely that the small difference between 
$\omega_1$ and $\omega_2$, which becomes much smaller at larger sizes, 
can be resolved within the density matrix renormalization group (DMRG),
we believe that at present the only possible approach is a semi-analytical 
one, e.~g., based upon symmetry-adapted trial wave functions for the 
lowest states $\Psi_{0,1,2}$, which also turned out useful for other strongly 
correlated electron systems.\cite{Baldea:2001b}

To end this section, we believe, in spite of the above somewhat speculative considerations, 
that one can plausibly ascribe the excitation energy $\omega_1$ as the width of the 
Kondo resonance, while the excitation energy $\omega_2$, close to but still 
different from $\omega_1$,
can be interpreted as the splitting of the Kondo resonance observed  
experimentally.\cite{Kogan:04a}
\section{FIR absorption}
\label{sec-fir}
In this section we shall focus 
on the other two transitions $\vert \Psi_0 \rangle \to \vert \Psi_{3,4} \rangle $.
As seen in  Fig.\ \ref{fig:abs}a, the absorption frequencies $\omega_{3,4}$
are of the order of $U$. For many fabricated SETs (see, \emph{e.~g.}, 
Refs.\ \onlinecite{Goldhaber-GordonNature:98}, \onlinecite{Goldhaber-GordonPRL:98}, and 
\onlinecite{liu:08}) these values belong to the FIR range.
The explicit forms (\ref {eq-g_u}) and (\ref{eq-Psi}) show that 
in the Kondo regime these two transitions amount to excite the electron from
the QD lower Hubbard band into the electrode Fermi level, and from the 
electrode Fermi level into the QD upper Hubbard band; 
sufficiently away from the mixed valence ranges ($\varepsilon_d \approx \varepsilon_F$,
$\varepsilon_d \approx \varepsilon_F - U$),
the exact excitation energies are well approximated by 
$\omega_3^0 = \varepsilon_F - \varepsilon_d $ and 
$\omega_4^0 = \varepsilon_d + U - \varepsilon_F $
(see Fig.~\ref{fig:abs}a). 

Based on Fig.\ \ref{fig:abs}, 
one expects in general two absorption peaks of a SET irradiated with 
FIR radiation. In the middle of the Kondo plateau 
($\varepsilon_d^\ast = \varepsilon_F - U/2$) the two transitions 3 and 4 
are degenerate, and therefore a single peak can be observed experimentally.
There, the absorption frequency is just one half of the charging energy,
$\omega_3 = \omega_4 = \omega^\ast \equiv U/2$.
By moving away from this point in either direction, the absorption peak 
splits into two peaks of different intensities 
located symmetrically with respect to the degenerate peak, 
$\omega_{3,4} \simeq \omega^\ast \mp \vert \varepsilon_d - \varepsilon_d^\ast\vert$. 
The farther from the symmetric point, the more pronounced is 
the asymmetry in intensity,
the stronger is the peak $\mu_3$ at the lower frequency $\omega_3$,
and the weaker the peak $\mu_4$ at the higher frequency $\omega_4$. 
\par
Out of the studies on SETs in ac fields,\cite{CampoOliviera:03,Sindel:05,Laakso:08}
excepting in part for Ref.~\onlinecite{CampoOliviera:03}, none considered
the above aspects. Without establishing any relationship
to the FIR absorption, the numerical results on frequency-dependent 
conductance deduced within the NRG 
of Ref.~\onlinecite{CampoOliviera:03} show, interestingly,  
a weak peak (to which the authors paid little attention) 
for two values of  $\varepsilon_d$: 
at $\varepsilon_d = \varepsilon_F - U$ and 
at $\varepsilon_d = \varepsilon_F - U/2$ (see Figs.~2 and 3, 
respectively of Ref.~\onlinecite{CampoOliviera:03}, to which we refer below). 
This peak is directly related to our results. The situation
$\varepsilon_d = \varepsilon_F - U/2$
corresponds just to the point of particle-hole symmetry, and the peak 
position is visible, just as predicted by the present approach, 
at $\omega = \omega^\ast $ (note that $\varepsilon_F $ 
is set to zero in Ref.~\onlinecite{CampoOliviera:03}).
For $\varepsilon_d = \varepsilon_F - U$, the peak in Fig.~2 of 
Ref.~\onlinecite{CampoOliviera:03} occurs at 
$\omega \sim (6\times 10^{-3}/0.025)\ U = 0.24 U$, but the authors 
provide no physical interpretation of this value. In excellent agreement 
with this value, the lower frequency absorption peak $\mu_3$ predicted 
by our approach is $\omega_3 = U/4$. In addition, we predict
another absorption peak $\mu_4$ at a higher frequency $\omega_4 = (3/4)U$
($\omega_4/D = 0.01875 $ in the notation of 
Ref.~\onlinecite{CampoOliviera:03}), which, although in the range 
showed in Fig.~3, is invisible there. We can explain this fact:
for the parameters employed in Ref.~\onlinecite{CampoOliviera:03},
\cite{U-too-large} we estimate that the higher frequency peak 
would be one order of magnitude less intense than the lower frequency one. 
This weak intensity could hardly be distinguished 
in the background of the curve of Fig.~3 at $\omega/D  = 0.01875$.
To reveal the two peaks in FIR absorption, the 
NRG calculations should have used situations sufficiently away from the 
particle-hole symmetry point ($ \vert \varepsilon_d - \varepsilon_d^\ast\vert$ 
should exceed the peak widths) but still sufficiently close to it, 
because otherwise the high frequency peak would be too weak and thence 
not visible. 

To end this section, we note that 
the two peaks in the FIR absorption at $\omega_3$ and $\omega_4$ are the counterparts of 
two maxima located close to the energies 
$\varepsilon_d$ and $\varepsilon_d + U$, which are present in 
the electronic density of states (DOS) along with the sharp peak corresponding to the Kondo resonance
(see, \emph{e.g.}, Fig.~3 of Ref. \onlinecite{swirkowicz:03}).
\section{Experimental implications}
\label{sec-exp}
Based on the above theoretical results, we propose to employ the 
FIR absorption as an experimental tool to characterize SETs. 
To avoid misunderstandings, we 
emphasize that the proposed experiments are different both from those 
carried out using rf or microwave radiation suitable for studying the 
Kondo resonance (e.~g, Ref.~\onlinecite{Kogan:04a})
and from those recently proposed by us to use photoionization,\cite{Baldea:2009a} 
where photons should have energies of the order of the work function 
(ultraviolet radiation). 
\par
In experiments, even using a very well focused flux of FIR 
photons to irradiate a SET, it is important but, fortunately, easy to 
discriminate between absorption processes occurring 
in the dot, and in electrodes or due to acoustic phonons. 
One should simply monitor 
absorption by varying $V_g$: the former signals are affected and should be analyzed, 
while the latter are not and should be disregarded. 
To exploit the present results,
most desirable would be to record FIR absorption spectra of SETs directly.
The absorption intensities may be very weak and their measurement 
a challenge for experimentalists. Even though difficult, this can no longer be considered 
a hopeless experimental task, particularly in view of the very recent advances in the field of 
molecular devices, enabling to measure the photon emission 
\cite{FluorescenceSingleMoleculeGalperin:2008}
or Raman response \cite{RamanConductionSingleMoleculeGalperin} of a \emph{single} molecule. 

As an easier experimental task, similarly to our earlier work \cite{Baldea:2009a}, 
we propose to perform a mixed FIR-absorption--dc-transport study, 
which should not pose special experimental problems. 
Again, the fact that in single molecules experimentalists were able to measure 
electronic conduction \emph{simultaneously} with the 
photon emission \cite{FluorescenceSingleMoleculeGalperin:2008} 
or Raman response \cite{RamanConductionSingleMoleculeGalperin} is very encouraging 
for the present proposal.
What one should monitor is the current $I$ at $T<T_K$
with an applied small dc source-drain voltage 
and subject to a monochromatic FIR radiation with tunable frequency.
To anticipate, most important for this experiment is that the 
absorption intensities need not be measured.
The manner to conduct the experiment
and to deduce the relevant parameters can easily be understood
by inspecting Fig.~\ref{fig:exp}. 
\par
Let us assume that the gate potential $V_g$ is increased, starting from a 
sufficiently negative value ($V_g  < V_{g,l}$), where the dot level is empty, 
and $I=0$. 
As soon as the Kondo plateau is reached ($V_g  \agt V_{g,l}$, 
$\varepsilon_d \alt \varepsilon_F $), which is signaled 
by the onset of a current flow $I \neq 0$,
FIR absorption becomes possible by appropriately tuning the  
frequency $\omega$ of incoming photons, $\omega = \omega_3(V_g)$. 
For ascertaining the resonance, it is not necessary 
to detect a nonvanishing absorption intensity: 
the resonance will be signaled by the current drop ($I\approx 0$),
which should be observable in a time-resolved dc-transport measurement, 
because, by absorbing a photon, the unpaired electron of the dot 
will be displaced into electrodes, and 
the prerequisite for the Kondo effect will disappear.
In this region, the second signal at the 
higher frequency $\omega_4$ could hardly be detected, because of its 
small intensity (see Fig.~\ref{fig:abs}b), and this is visualized by the 
dashed line in Fig.~\ref{fig:exp}. By further increasing $V_g$, it will acquire 
sufficient intensity; the representation in Fig.~\ref{fig:exp}
switches from a dashed to a solid line. There, a current drop is observed 
by tuning the photon frequency both to $\omega = \omega_{3}(V_g)$ and to 
$\omega = \omega_{4}(V_g)$.
The frequencies $\omega_3$ and $\omega_4$, which vary linearly with $V_g$, 
become closer and closer, and the two absorption signals tend to 
coalesce. Their overlap is perfect ($\omega_3 = \omega_4$)
at $V_g = V_g^\ast$ at the point of particle-hole symmetry. 
Beyond this point, 
the trend reverses: the lower frequency $\omega_3$ decreases 
while the upper frequency $\omega_4$ increases, and the latter
signal eventually becomes too weak to produce a current drop 
(the line switches from solid to dashed).
\par
Importantly, $U$ and $\alpha$ can be determined from the curves $\omega_{3,4}(V_g)$.
The former can be deduced from the location of 
the two overlapping absorption peaks in the middle of the Kondo plateau,
$U = 2 \omega_{3,4}(V_g^\ast)$. If there were uncertainties 
to exactly locate the point of the perfect overlapping, 
one could alternatively use the intersection of 
the extrapolated $\omega_3$- and $\omega_4$-lines.
This is a direct determination of $U$, and not an indirect one, 
as in low-bias dc-measurements, for which the conversion factor $\alpha$ is needed.
Moreover, even $\alpha$ can be directly obtained from the slope of the curves, 
$\alpha = \mp d\omega_{3,4}/d V_g$, or, alternatively, by using the value 
of $U$ [$\alpha = U/(V_{g,l} - V_{g,u})$]. So, one can even perform a 
self-consistency test. Once an accurate $\alpha$-value is available, 
one can use the extension $\delta V_g$ of the Kondo plateau edges
to obtain the parameter $\Gamma \sim t_d^2/D$, which characterizes 
the finite level width induced by the QD-electrode coupling. 
This is also important, because $t_d$ can also be controlled experimentally
by varying the gate potentials that form the constrictions. 
\cite{Goldhaber-GordonPRL:98}
\begin{figure}[htb]
\includegraphics[width=0.33\textwidth,angle=-90]{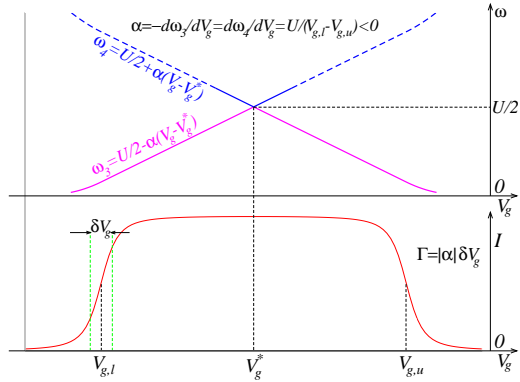}
\caption{\label{fig:exp} (Color online) Schematic representation of 
the change in the FIR absorption frequencies $\omega_{3,4}$ and dc current $I$
by varying the gate potential $V_g$, which permits to deduce the 
parameters $U$, $\alpha$, and $\Gamma$. 
See the main text for details.}
\end{figure}
\section{Conclusion}
\label{sec-conclusion}
The smaller the size of a quantum dot, the larger is its charging energy.
Larger QDs possess smaller charging energies (for example, 
$U \approx 64\,\mu$eV \cite{huang:08}),
and therefore they could be investigated by rf or microwave techniques.
However, smaller QDs, as those often used in a SET setup, are characterized by 
considerably larger charging energies (\emph{e.~g.},
$U \approx 1.9$\,meV,\cite{Goldhaber-GordonPRL:98} or $U \approx 7-8$\,meV \cite{liu:08}),
and, consequently, for them the aforementioned techniques cannot be directly employed.
In the present paper, we have presented theoretical results demonstrating 
that FIR experiments 
on such SETs, which are feasible, permit to accurately determine
the charging energy and other important parameters
in a direct way. Concerning the FIR absorption, three aspects are worth to be mentioned. 

First, we emphasize that, 
in comparison with other methods, the FIR absorption possesses important advantages.
It is not affected by parasitic currents due to unavoidable capacitive couplings,
as it is the case of rf or microwave techniques. Likewise, it is much less challenging 
than photoionization studied recently:\cite{Baldea:2009a} in the FIR experiments 
discussed in the 
present paper one simply needs to determine absorption energies of the order 
of a few meV with a reasonable accuracy, while photoionization requires 
the determination of ionization energies (of the order of the work functions, 
typically $\sim 1$\,eV) with an accuracy $\sim 1$\,meV.\cite{Baldea:2009a} 

Second, we note that the investigation with the aid of FIR radiation is by no means 
limited to SETs. In nanodevices based on double (or other assembled) QDs, FIR absorption 
can also be used to deduce other relevant parameters,\cite{Baldea:unpublished} like the 
interdot electrostatic coupling (or $V$-Hubbard strength), which are related to 
important properties of nanostructures (see, \emph{e.g.}, Refs.\ \onlinecite{Baldea:2008} and 
\onlinecite{Baldea:2009b}), 
and which cannot be straightforwardly deduced from zero-bias dc-conductance data.

Third, one should emphasize the 
cross-fertilization between NRG and exact numerical diagonalization.
Based on a few significant many-body configurations, the latter method 
is very intuitive and allowed us to give a physical content to the NRG numerical 
findings unraveled so far. Conversely, the agreement 
between the NRG results, valid for infinite electrodes, and the exact 
diagonalization, which can be carried out only for short electrodes, demonstrates
that the latter is able to make certain valuable predictions that are not affected by 
finite-size effects, as already noted.\cite{Baldea:2008b,Baldea:2009a}

In addition to the FIR absorption, in the present paper we presented results 
on the SET microwave/rf absorption, which, although preliminary, are interesting 
in the context of the recent experiments revealing the splitting of the 
Kondo resonance.\cite{Kogan:04a,Wingreen:04}  
We hope to return soon to this important issue, which deserves further 
work.
\section*{Acknowledgments}
I.~B.~is indebted to M.~Galperin for bringing 
Refs.~\onlinecite{FluorescenceSingleMoleculeGalperin:2008} and
\onlinecite{RamanConductionSingleMoleculeGalperin} to his attention. 
The authors acknowledge with thanks the financial support for this work 
provided by the Deu\-tsche For\-schungs\-ge\-mein\-schaft (DFG).

\begin{thebibliography}{10}
\bibitem{1ph-2ph}
In Figs.~\ref{fig:abs}a and \ref{fig:all-omega}a, the sudden doubling in the
  slopes of the $\omega_3$- and $\omega_4$- curves around the mixed-valence
  points of coordinates $(-4, 8)$ and $(4, 8)$, respectively result from two
  avoiding crossings between the curves of one-particle--one-hole and
  two-particle--two-hole excitations mentioned in the text. In either case,
  only one mate of the two states involved in the avoided crossing is shown.
  The other mate, which is many orders of magnitude less intense, can be
  reconstructed by following the thin lines of Fig.~\ref{fig:all-omega}a.
\bibitem{U-too-large}
For a bandwidth $D=2$\,eV, the charging energies in Figs.~2 and 3 of
  Ref.~\onlinecite{CampoOliviera:03} are $U=0.025D=50$\,meV and
  $U=0.05D=100$\,meV, respectively, and thus roughly one or two orders of magnitude
  larger than for real SETs.
\end{thebibliography}

\end{document}